\documentclass[english,aps,pre,twocolumn,showpacs,superscriptaddress,floatfix]{revtex4-1}
\usepackage[T1]{fontenc}
\usepackage[latin1]{inputenc}
\usepackage{amsmath}
\usepackage{graphicx}
\usepackage{amssymb}
\usepackage{dcolumn} 
\usepackage{bm} 
\usepackage{color}

\makeatletter

\usepackage{amsfonts}
\usepackage{epsfig}

\newcommand{\Eq}[1]{Eq.~(\ref{#1})}

\newcommand{\be}{\begin{equation}}
\newcommand{\bea}{\begin{eqnarray}}
\newcommand{\eea}{\end{eqnarray}}
\newcommand{\ee}{\end{equation}}

\newcommand{\quot}[1]{``#1''}

\usepackage{babel}
\makeatother

\begin{document}

\title{Fractional Gaussian noise criterion for correlations characterization: a random-matrix-theory inspired perspective}

\author{Tayeb \surname {Jamali}}
\affiliation{Department of Physics, Shahid Beheshti University, G.C., Evin, Tehran 19839, Iran}

\author{Hamed \surname {Saberi}}
\affiliation{Department of Physics, Shahid Beheshti University, G.C., Evin, Tehran 19839, Iran}
\affiliation{School of Physics, Institute for Research in Fundamental Sciences (IPM), Tehran 19395-5531, Iran}

\author{G. R. \surname {Jafari}}
\affiliation{Department of Physics, Shahid Beheshti University, G.C., Evin, Tehran 19839, Iran}
\affiliation{School of NanoScience, Institute for Research in Fundamental Sciences (IPM), Tehran 19395-5531, Iran}

\date{May 26, 2013}

\begin{abstract}

We introduce a particular construction of an autocorrelation matrix
of a time series and its analysis based on the random-matrix theory
ideas that is capable of unveiling the type of correlations
information which is inaccessible to the straight analysis of the
autocorrelation function. Exploiting the well-studied hierarchy of
the fractional Gaussian noise (fGn), an \emph{in situ} criterion for
the sake of a quantitative comparison with the autocorrelation data
is offered. We illustrate the applicability of our method by two
paradigmatic examples from the orthodox context of the stock markets
and the turbulence. Quite strikingly, a remarkable agreement with
the fGn is achieved notwithstanding the non-Gaussianity in returns
of the stock market. In the latter context, on the contrary, a
significant deviation from an fGn is observed despite a Gaussian
distribution of the velocity profile of the turbulence.

\end{abstract}

\pacs{05.40.-a; 05.45.Tp}


\maketitle

\section{Introduction}
\label{sec:intro}

Physical entities typically loose their individual identities within a highly correlated whole which is not describable in terms of the mere knowledge of its physical constituents. Viewing a time series as a whole system with the series values as its constituents, the corresponding autocorrelation can be interpreted as the correlations among the system's constituents. The random-matrix theory (RMT) is the method of choice for identifying such correlations~\cite{Laloux, Plerou1, Plerou2, Plerou3}. However, a blind comparison of such correlation information with the statistical characteristics of the relevant RMT \emph{per se} reveals only the discrepancy from a purely random matrix statistics without relying on an \emph{a priori} physical knowledge about the system. The fractional Gaussian noise (fGn)~\cite{Mandelbrot}, though, provides an appropriate measure to trace back systematically the fingerprints of such physical information left already on the correlation profile of the system. Some recent considerations of such a figure of merit include invoking an fGn-based criterion for the visibility method by Lacasa \emph{et al}.~\cite{Lacasa} and an fGn analysis of the level crossing algorithm by Vahabi \emph{et al}.~\cite{Vahabi}.

FGns are known, well-studied, and describable by a single Hurst exponent. This promises the following proposed strategy for keeping track of the correlations: One first calculates the autocorrelation matrix of the empirical series and then rather than comparing the outcome with the Hurst exponent of a white noise (known to be $0.5$) in RMT, a detailed comparison is being made with the whole family of the Hurst exponents of the fGns. In this way, the \emph{optimal} Hurst exponent yielding the most consistent interpretation of those empirical series can be extracted. In practice, a direct comparison of the ensemble average of the eigenvalues spectra of the time series associated with various values of $H$ with that of the original time series can provide such an optimal Hurst exponent. It is the purpose of the present paper to demonstrate such ideas.

This paper is structured as follows: Sec.~\ref{sec:methodology} sets the scene by providing the details of the main technique introduced for an autocorrelation analysis of a time series as an alternative to RMT. The statistics of the fGn series based on such an autocorrelation analysis is next detailed in Sec.~\ref{sec:Statistics of fGn}. Sec.~\ref{sec:real_life} illustrates the application of the proposed method through two examples in the context of the finance and turbulence. Finally, Sec.~\ref{sec:conclusions} contains our conclusions and an assessment of the applicability range of the proposed formalism in other relevant contexts.

\section{Methodology}
\label{sec:methodology}

\subsection{Random-matrix theory}
\label{subsec:RMT}

RMT is a widely used mathematical technique for studying the statistical characteristics of large complex systems where the nature of the underlying interactions and their associations to the ensuing correlations is not known \emph{ab initio}. For instance, the experimental data from nuclear scattering exhibit a great deal of complexity associated with the stochastic behavior of the nuclear resonances and their spacings~\cite{Porter}. Eugene Wigner used RMT to describe the statistical distribution of such nuclear resonances~\cite{Wigner1} and found a striking resemblance between such a distribution and the one associated with the eigenvalues of a known class of random matrices called Gaussian orthogonal ensemble (GOE). RMT has since then found a wide range of applications in various and seemingly unconnected areas of mathematics and physics including the number theory~\cite{Schroeder}, finance~\cite{Laloux, Plerou1, Plerou2, Plerou3, Namaki1, Namaki2, Namaki3}, and quantum many-body systems~\cite{Guhr}. Such an extensive list of applications is believed to be rooted in the existence of the \emph{universality classes} which, in turn, is conceivably a consequence of the underlying symmetries and the law of large numbers~\cite{Ergun}. Here, we provide only a brief overview of the method at the minimal level on which the subsequent materials rely and refer instead the interested reader to Refs.~\cite{Brody, Ergun, Guhr, Mehta} for further and deeper details.

Random matrices are matrices with random elements which are drawn from some probability distribution subject to some required symmetries. Dyson already demonstrated that all random matrix ensembles do fall into three universality classes called the orthogonal ensemble (OE), unitary ensemble (UE) and symplectic ensemble (SE)~\cite{Dyson1,Dyson2,Dyson3}. The universality class associated with a given random matrix ensemble is determined by the transformation-invariance properties of its distribution function and the type of physical quantity (e.g. the Hamiltonian, scattering or transfer matrix) the random ensemble represents. As an instance, consider an $M \times M$ Hamiltonian matrix $\mathbf{H}$ with \emph{independent} matrix elements $H_{ij}$ taken from some random distribution $P_{ij}$ and, whereupon, a total distribution function of the product form $P(\mathbf{H}) \equiv \prod_{ij} P_{ij}(H_{ij})$. Then the invariance of the distribution function of the Hamiltonian $P(\mathbf{H)}$ so defined under each orthogonal, unitary, or symplectic transformation specifies its functional form as a Gaussian distribution of the form
\begin{eqnarray}
\label{eq:P(H)}
P_{M\beta}(\mathbf{H}) = c_{M\beta} \exp\bigg(-\frac{M\beta}{4\sigma^2}\textrm{Tr}\{\mathbf{H}^2\}\bigg) \; ,
\end{eqnarray}
where $c_{M\beta}$ is the  normalization constant, $\sigma$ denotes the standard deviation of the off-diagonal matrix elements, and the cases $\beta= 1, 2, 4$ correspond to a Gaussian orthogonal ensemble (GOE), a Gaussian unitary ensemble (GUE), and a Gaussian symplectic ensembles (GSE), respectively.

The joint probability distribution of the eigenvalues of the random matrix $\mathbf{H}$ (denoted by $\lambda_1,...,\lambda_M$) for all Gaussian ensembles can be calculated from \Eq{eq:P(H)} and is given by
\begin{multline}
\label{eq:P(l)}
P_{M\beta}(\lambda_1,...,\lambda_M) \\ = c_{M\beta} \prod_{i<j} |\lambda_i-\lambda_j|^\beta \exp\bigg(-\frac{M\beta}{4\sigma^2} \sum_{k=1}^M \lambda_k^2\bigg) \;,
\end{multline}
where the dependence $|\lambda_i-\lambda_j|^\beta$ indicates the repulsion of the adjacent eigenvalues. For the specific case of $2 \times 2$ matrices, the distribution of the spacings between the nearest-neighbor eigenvalues $s \equiv \Delta \lambda$ can be obtained from the latter equation and is known to be given by~\cite{Ergun}
\begin{eqnarray}
\label{eq:P(s)}
P_{\beta}(s) = c_{\beta} s^\beta \exp\big(-a_{\beta}s^2 \big) \; ,
\end{eqnarray}
where $c_{\beta}$ and $a_{\beta}$ are some $\beta$-dependent constants. The relation is often referred to as the Wigner surmise~\cite{Wigner3} and can be shown to be yet valid to a good approximation even in the limit $M \rightarrow \infty$ ~\cite{Gaudin}.

\subsection{Fractional Gaussian noise}
\label{subsec:fGn}
FGns arise naturally as a by-product of the idea of fractional Brownian motion (fBm) introduced originally by Kolmogorov~\cite{Kolmogorov} as a generalization of the ordinary Brownian motion (Bm). Further developments were made by Mandelbrot and van Ness~\cite{Mandelbrot} who proposed a stochastic integral representation of fBm as a continuous-time integrator $B_{H}$ of the form
\begin{multline}
\label{eq:fBm}
B_H(t)=\frac{1}{\Gamma(H+1/2)} \bigg(\int_{-\infty}^{0} \big[(t-\theta)^{H-1/2} \\ -(-\theta)^{H-1/2}\big]dB(\theta)+\int_0^t(t-\theta)^{H-1/2}dB(\theta)\bigg) \; ,
\end{multline}
where $\Gamma(H)$ is the usual Gamma function and $0<H<1$ denotes the Hurst exponent. Note that for $H=0.5$ the ordinary Bm is recovered. Just as the Bm, fBm is also a Gaussian process and belongs to the zero-mean class of the stochastic processes~\cite{Mandelbrot}. The only difference between them, though, lies in their autocovariance obtainable from the latter equation and given by~\cite{Mandelbrot}
\begin{eqnarray}
\label{eq:autocovariance}
\langle B_H(t)\;B_H(t')\rangle_{\mathrm{ens.}}=\frac{c_H}{2} \big[{t'}^{2H}+t^{2H}-(t'-t)^{2H}\big]  \; ,
\end{eqnarray}
where $\langle\cdots\rangle_{\mathrm{ens.}}$ denotes the ensemble average, $0 < t \le t'$, and the coefficient $c_H$ is defined through~\cite{Barton}
\begin{eqnarray}
\label{eq:c_H}
c_H \equiv \Gamma(1-2H)\cos(\pi H)/\pi H \; .
\end{eqnarray}
As is easily seen from the autocovariane expression (\ref{eq:autocovariance}), fBm is not stationary and its standard deviation varies with time albeit is endowed with stationary increments~\cite{Bradley}.

Moreover, it is known that the distribution functions associated with a Gaussian process are uniquely determined from the knowledge of the mean and autocovariance structures~\cite{Feller}. Therefore, given that $B_H (at)$ and $|a|^H B_H (t)$ (with $a$ as an arbitrary parameter) have equal values of mean and autocovariance, we can deduce that they feature the same distributions. This, in turn, implies the self-similarity of the fBm. Further details of the fGns can be found in Ref.~\cite{Bradley} and the references therein.

The derivative of fBm, on the other hand, yields the so-called fGn~\cite{Bradley}. Although such a derivative does not exist in a rigorous mathematical sense, but nevertheless an fGn can be defined through the discrete increments of fBm as
\begin{eqnarray}
\label{eq:G_H}
G_H(k)\equiv B_H(k+1)-B_H(k) \quad \mathrm{for} \quad k\geq1 \; .
\end{eqnarray}
The quantity so defined has a normal distribution for every integral input parameter $k$, but exhibits a long-range dependence save for the case of $H=0.5$ associated with the ordinary Bm. Its autocovariance for integral values of $n$ also takes the form
\begin{multline}
\label{eq:auto_cov_fBm}
\langle G_H(k)\;G_H(k+n)\rangle_{\mathrm{ens.}} = \frac{c_H}{2} \big[|n-1|^{2H}-2|n|^{2H} \\ +|n+1|^{2H}\big] \; .
\end{multline}

It follows then that fGn is a stationary Gaussian process with the type of the underlying correlation determined from the sign of the corresponding autocovariance. Depending on the value of the Hurst exponent, three regimes are identifiable:

(i) For $H= 0.5$ there is no correlation and the sample mimics the white noise.

(ii) For $H < 0.5$ the noise is \emph{negatively} correlated and the sample fluctuates faster than the white noise.

(iii) For $H > 0.5$ the noise is \emph{positively} correlated and the sample fluctuates slower than the white noise.

It is noteworthy to mention that the autocovariance of fGn behaves asymptotically as $c_H H (2H-1)n^{2H-2}$ with a long-range dependence for $0.5<H<1$ and tends to zero as $n \to \infty$, implying the ergodicity of fGns~\cite{Samorodnitsky,Sinai}. Furthermore, fGn features self-similarity just as the fBm.
In this work, a wavelet-based simulation~\cite{Dieker} has been utilized as an approximation method for generating fGn series.

\subsection{Autocorrelation matrix of a time series}
\label{subsec:autocorr_fGn}

Quantification of the correlations among the system's constituents is an issue of central importance in all areas of physics and has been the subject of intensive scientific investigations. Basic insight can be gained into the nature of such correlations by diagonalizing the so-called \emph{correlation matrix} whose elements quantify the way the system's constituents may affect each other. The so-obtained statistics of the eigenvalues and eigenvectors is compared with that of a random matrix where deviations from which characterize the desired correlation information~\cite{Plerou3}.

In this work, however, we propose a different approach according to which one first treats an empirical time series as the \emph{whole} system and the series values as its constituents. As such, the corresponding autocorrelation can be interpreted as the correlations among the system's constituents. We propose, furthermore, fGns as the figure of merit to interpret the data from the target empirical series by a comparison between two in terms of the statistics of their autocorrelation matrix. The statistical comparison is performed via studying the dependence of three major characteristics of the autocorrelation matrix of fGns on the Hurst exponent $H$, namely the distribution of the eigenvalues, the distribution of the nearest-neighbor spacings between the eigenvalues, and the number of significant participants in an eigenvector, which shall be all detailed in Sec.~\ref{sec:Statistics of fGn}. We point out our proposed fGn criterion is inspired by the RMT in considering such statistical figures of merit heavily exploited in the latter theory~\cite{Laloux, Plerou1, Plerou2, Plerou3}.

The symmetric \emph{autocorrelation matrix} $\mathbf{C}$ of a time series $X=\{X_t:t=1,\ldots,T\}$ of length $T$ is given through its matrix elements
\begin{eqnarray}
\label{eq:auto_corr_C}
C_{t,t+\bigtriangleup t} =\frac{\langle X_t X_{t+\Delta t}\rangle_{\mathrm{time}}-\langle X_t\rangle_{\mathrm{time}} \langle X_{t+\Delta t}\rangle_{\mathrm{time}}}{\sigma^2} \; ,
\end{eqnarray}
where $\sigma = \sqrt{\langle X^2\rangle_{\mathrm{time}} -\langle X\rangle_{\mathrm{time}}^2}$ is the standard deviation of $X_t$, and $\langle \ldots\rangle_{\mathrm{time}}$ denotes the time average over the period of the series. Note that the time lag $\Delta t$ ranges from $1-t$ to $N-t$ to assure the construction of an $N \times N$ autocorrelation matrix for every $t \in [1,N]$.

In order to realize an unbiased and uniform comparison between fGns with different Hurst exponents, their mean and variance are set to 0 and 1, respectively. Note that the latter amounts to setting the coefficient $c_H$ in \Eq{eq:auto_cov_fBm} to unity. We can establish then the autocorrelation matrix and investigate its eigenvalues spectrum (ES), the distribution of nearest-neighbor eigenvalues spacings (NNES), and inverse participation ratio (IPR). It must be noted, however, that such a recipe is prone to introduce two major size effects into the numerical calculations as follows: The finite length $T$ of the time series $X$ and the finite size $N$ of the autocorrelation matrix $\mathbf{C}$. Among the two, the former finite size effect associated with the finite length of the time series can nevertheless be removed by construction thanks to the ergodicity feature of fGns~\cite{Samorodnitsky,Sinai} which makes the ensemble average in \Eq{eq:auto_cov_fBm} to be equal to the time average in \Eq{eq:auto_corr_C}. This, in turn, allows one to read the autocorrelation matrix $\mathbf{C}$ directly from the expression of \Eq{eq:auto_cov_fBm}. We call the autocorrelation matrix so obtained the \emph{length-free autocorrelation matrix} (LFAM) $\mathbf{\tilde{C}}$ whose matrix elements may be calculated from
\begin{multline}
\label{eq:length-free_auto_cov}
\tilde{C}_{t, t+\Delta t} = \frac{1}{2} \big[|\Delta t-1|^{2H}-2|\Delta t|^{2H} \\ +|\Delta t+ 1|^{2H}\big] \; ,
\end{multline}
and provide in the subsequent section the results of the calculation of its ES, the distribution of NNES, and IPR for an illustrative value of $N$ and finally the finite-size effect associated with the finite size $N$ of such an autocorrelation matrix $\mathbf{\tilde{C}}$.

\section{Statistics of the fractional Gaussian noise} 
\label{sec:Statistics of fGn}

\subsection{Eigenvalues distribution of the length-free autocorrelation matrix}
\label{subsec:Eigenvalue_dist}
As an illustrative case, the ES of the LFAM with $N= 2000$ for various values of the Hurst exponents $H$ is shown in Fig.~\ref{fig:ES}. First of all, it is seen that the eigenvalues $\lambda_i$ ($i = 1 ,\cdots, N$) are positive for all values of $H$. Besides, three behavior regimes for the spectrum depending on the value of $H$ may be identified:

\begin{figure}[tb]
\centering
\includegraphics[width=1\linewidth]{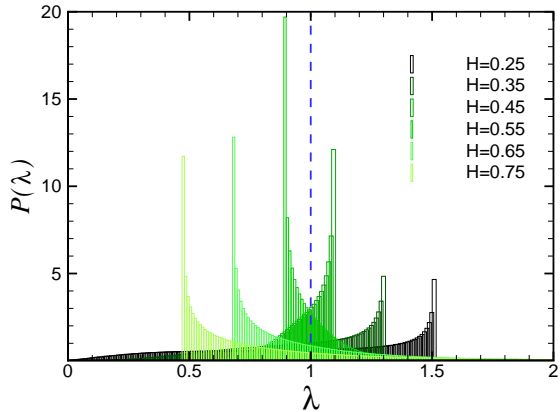}
\caption[The eigenvalues spectrum]{(Color online) The ES of the LFAM $\mathbf{\tilde{C}}$ of the size $2000\times2000$ for various values of the Hurst exponent $H$. The probability distribution function $P(\lambda)$ gives the abundance of a particular eigenvalue $\lambda$ in the spectrum or equivalently the quantity $P(\lambda) \Delta\lambda$ can be interpreted as the probability that an eigenvalue of $\mathbf{\tilde{C}}$ arises in the interval $[\lambda, \lambda + \Delta\lambda]$. The vertical dashed line also corresponds to the LFAM with a Hurst exponent $H=0.5$ for which the distribution function exhibits a Dirac-delta like singularity indicating that all the eigenvalues are equal to unity.
}
\label{fig:ES}
\end{figure}

 (i) For $H<0.5$ the eigenvalue with the maximal abundance $\lambda_{\mathrm{max}}$ is the largest one and bounded from above.

 (ii) For $H= 0.5$ all eigenvalues equal unity giving rise to a Dirac delta function denoted by a dashed line in Fig.~\ref{fig:ES}.

 (iii) For $H>0.5$ the eigenvalue with the maximal abundance $\lambda_{\mathrm{max}}$ is the smallest one and bounded from below. Besides, the limiting value of $\lambda_{\mathrm{max}}$ turns out to be unity upon reaching the value of $H=0.5$ both from above and below.


\subsection{Distribution of the nearest-neighbor eigenvalues spacings for the length-free autocorrelation matrix}
\label{subsec:Eigenvalue_spacing}

\begin{figure}[tb]
\centering
\includegraphics[width=1\linewidth]{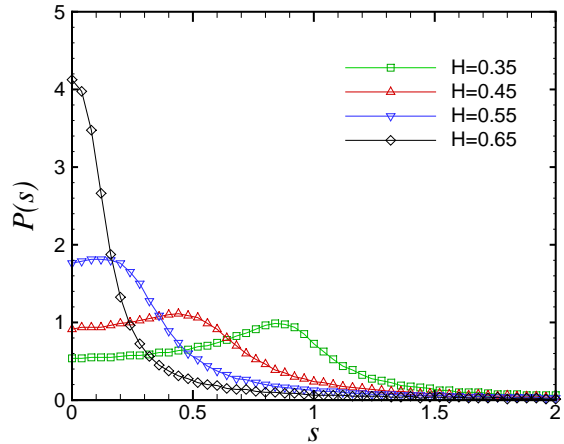}
\caption[Distribution of nearest-neighbor eigenvalues spacings]{(Color online) Distribution of the NNES of the LFAM of the size $2000 \times 2000$ for various values of the Hurst exponent $H$ while employing a Gaussian broadening recipe as the required unfolding transformation. The vertical axis represents the probability distribution function of NNES $P(s)$ as the abundance of a particular NNES $s$ in the spectrum or equivalently the quantity $P(s) \Delta s$ can be interpreted as the probability that an NNES of the value $s$ arises in the interval $[s, s + \Delta s]$ in the ES of the LFAM $\mathbf{\tilde{C}}$.}
\label{fig:NNES}
\end{figure}

We now aim at calculating the distribution of NNES of the LFAM for a given value of $N$. Here since the level spacing varies throughout the eigenvalues spectrum, we need an \quot{unfolding} procedure to transform the original eigenvalues $\lambda_i$ into properly rescaled and dimensionless ones $\tilde{\lambda_i}$ ~\cite{Mehta,Brody,Guhr}. More precisely, the unfolding procedure provides a local rescaling of the eigenvalues spectrum with respect to the local average of the level spacing. As a result of such a rescaling scheme, the local average of level density remains constant and independent of $\lambda_i$ and thereby comparable to those from RMT. As such, rather than the \quot{bare} distribution of the original eigenvalue spacings, i.e, $\lambda_{i+1}-\lambda_{i}$, that of the \emph{unfolded eigenvalues}, i.e., $s_i\equiv \tilde{\lambda}_{i+1}-\tilde{\lambda}_i$ is analyzed.

Figure~\ref{fig:NNES} illustrates the distribution of the unfolded NNES for the LFAM of the size $2000\times2000$ and for various values of the Hurst exponents while adopting a Gaussian broadening method~\cite{Bruus,Haake} for the sake of realizing the desired unfolding of the eigenvalues. It can be inferred from such a figure that the levels tend to each other on average (the NNES associated with the maximal value of $P(s)$ approaches zero) upon increasing $H$. They also end up closer to each other for $H> 0.5$ compared to the regime $H<0.5$.

\begin{figure}[tb]
\centering
\includegraphics[width=1\linewidth]{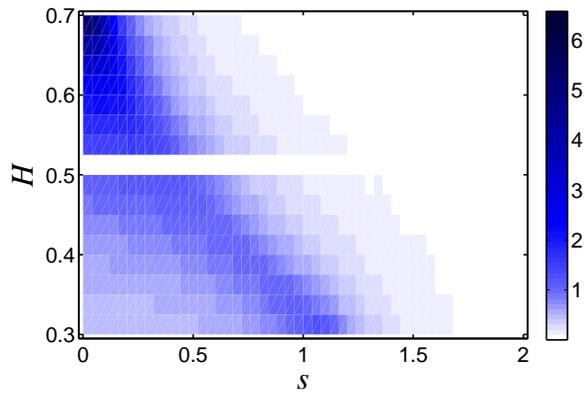}
\caption[Density plot]{Density plot of the distribution of NNES $P(s)$ for an interval of $H \in [0.3 , 0.7]$ on a color scale ranging between 0 and 6.5. The white ribbon in the middle of the vertical axis reflects the fully degenerate ES associated with the situation in which $H=0.5$ [compare to Fig.~\ref{fig:ES}]. The highest abundance of NNES associated with the peaks in Fig.~\ref{fig:NNES}, on the other hand, follow an almost diagonal trend of relatively narrow dark regions in the plot.}
\label{fig:NNES_total}
\end{figure}

Figure~\ref{fig:NNES_total} illustrates additionally the change in the distribution of NNES through a density plot for a finer spectrum of the values of $H$ compared to the previous figure.

\subsection{Inverse participation ratio of the eigenvectors of the length-free autocorrelation matrix}
\label{subsec:IPR}
We exploit here the notion of the inverse participation ratio (IPR) heavily arisen in the context of the theory of localization~\cite{Guhr} to determine the number of significant participants of each eigenvector given by
\begin{eqnarray}
\label{IPR}
I^k=\sum_{n=1}^N (u_n^k)^4 \;,
\end{eqnarray}
where $k = 1,\cdots, N$ and $u_n^k$ is the $n$'th component of the $k$'th eigenvector $\textbf{u}^k$. The number of significant components of an eigenvector is \emph{inversely} proportional to the value of the IPR so defined.

\begin{figure}[!b]
\centering
\includegraphics[width=1\linewidth]{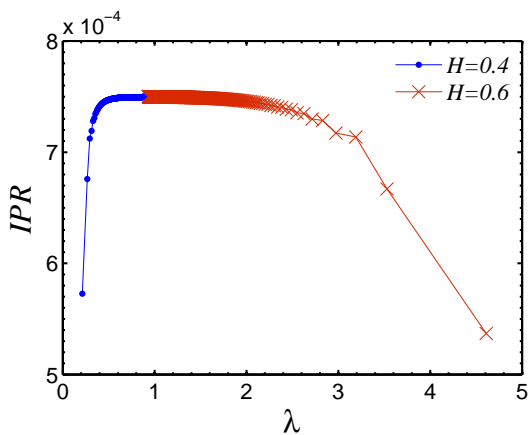}
\caption[IPR]{(Color online) The IPR of all eigenvectors of an LFAM of the size $2000\times2000$ for two illustrative values of $H=0.4$ and $H=0.6$.}
\label{IPR_total}
\end{figure}

Figure~\ref{IPR_total} depicts the IPR of all eigenvectors of the LFAM with respect to the associated eigenvalues $\lambda$ for two different Hurst exponents $0.4$ and $0.6$. As can be learnt from the plot, the IPR is an ascending function of $\lambda$ for $H=0.4$ and a descending one for $H=0.6$. We have checked numerically that the same trend continues with other numerically accessible values in two distinguishable regimes of $H<0.5$ and $H>0.5$, respectively. For $H = 0.5$, the IPR evidently remains constant.

\begin{figure}[tb]
\centering
\includegraphics[width=1\linewidth]{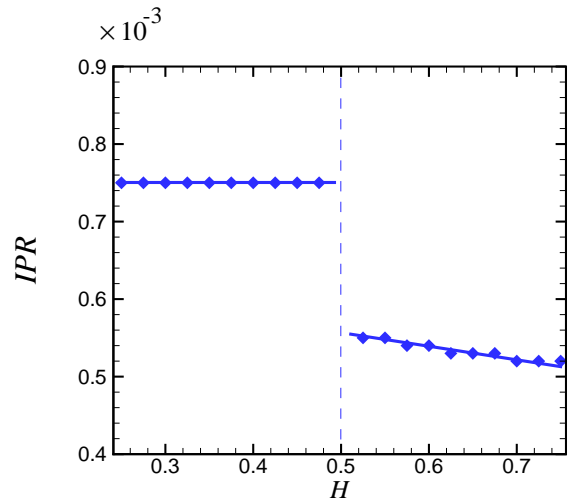}
\caption[???]{IPR of the eigenvectors associated with the largest eigenvalues of the LFAM plotted as a function of the Hurst exponent $H$. The vertical dashed line indicates the critical Hurst exponent that separates the negatively correlated regime from the positively correlated one associated with $H<0.5$ and $H>0.5$, respectively.}
\label{IPR}
\end{figure}

\begin{figure}[!b]
\centering
\includegraphics[width=1\linewidth]{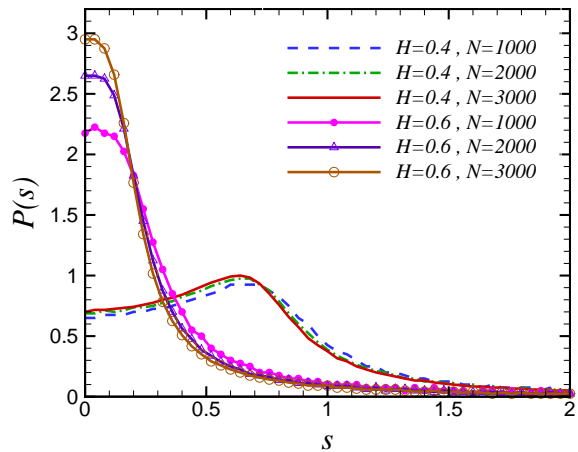}
\caption[Finite-size]{(Color online) Finite-size effect of the LFAM due the finiteness of its size $N$ on the distribution of NNES $P(s)$ for two values of the Hurst exponent $H$. To explore such an effect, for a fixed value of the Hurst exponent $H$, the distribution $P(s)$ has been plotted versus the NNES $s$ for three different values of the finite size $N$.}
\label{NNES_size_effect}
\end{figure}

The results for the IPR of the eigenvectors associated with the largest eigenvalues of a typical LFAM of the size $2000\times2000$ for different values of $H$ is additionally plotted in Fig.~\ref{IPR}. The result implies that the number of important participants of the eigenvector associated with the largest eigenvalue remains almost constant throughout the negatively correlated regime $(H<0.5)$ whereas it turns out to be relatively larger in the positively correlated region $(H>0.5)$ and rises upon increasing the Hurst exponent $H$.

\subsection{Finite size effect of the length-free autocorrelation matrix}
\label{subsec:size_effect}

Figure~\ref{NNES_size_effect} illustrates the effect of the finite size $N$ of the LFAM on the distribution of the NNES by providing the data for various values of the finite size $N$. As is evident in the figure, the size effect influences the data associated with $H>0.5$ much more significantly than those of $H<0.5$.

\section{Application to real-life time-series}
\label{sec:real_life}

\begin{figure}[t]
\centering
\includegraphics[width=1\linewidth]{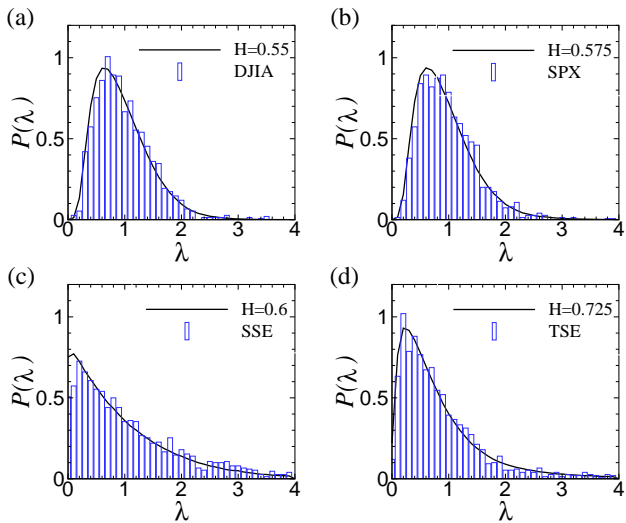}
\caption[ES_finance]{(Color online) The ES of the normalized daily return of DJIA, SPX, SSE, and TSE (the bars) contrasted to the ES of the optimal Hurst exponent of the associated fGn family (the solid line). A remarkable agreement with fGns is noticed upon such a comparison.}
\label{ES_finance}
\end{figure}

Finally, we demonstrate our ideas and in particular the relevance of the fGn criterion introduced in previous sections for the analysis of two paradigmatic real-life contexts, namely the finance time series and the phenomenon of turbulence. Before embarking on the application of the method to such examples, we first provide the details on how to obtain the optimal Hurst exponent associated with the searched member of the family of fGns that gives the most accurate description of the underlying correlations in a time series compared to the other exponents: To this end, one first produces ensembles of fGn series of the same finite length as the real one for various values of $H$. One proceeds by calculating the ES and NNES distribution of the ensemble average of the fGn series associated with such values of $H$ and compare eventually the outcome with that of the real time series. The comparison for a particular value of $H$ is made based on the calculation of the \emph{relative error} in ensemble average of the ESs with respect to that of the original time series. More precisely, given a time series of length $T$ and $N\times N$ autocorrelation matrix with eigenvalues $\mathbf{\Lambda}^0\equiv(\lambda_1^0, \lambda_2^0, \ldots\, \lambda_N^0)$, we generate $M$ fGns of size $T$ for each value of $H$ and construct $M$ autocorrelation matrices with eigenvalues $\mathbf{\Lambda}^m\equiv(\lambda_1^m, \lambda_2^m, \ldots,\lambda_N^m)$, $m = 1,2,\ldots,M$. Finally, for each Hurst exponent $H$, the two-norm relative error~\cite{Anderson} of ensemble average
$\overline{\mathbf{\Lambda}}=\frac{1}{M}\sum_{m=1}^{M}\mathbf{\Lambda}^m$ with respect to $\mathbf{\Lambda}^0$ is obtained. The optimal Hurst exponent corresponds then to the one leading to the smallest relative error.

\subsection{Financial time-series}
\label{sec:financial series}

\begin{figure}[!t]
\centering
\includegraphics[width=1\linewidth]{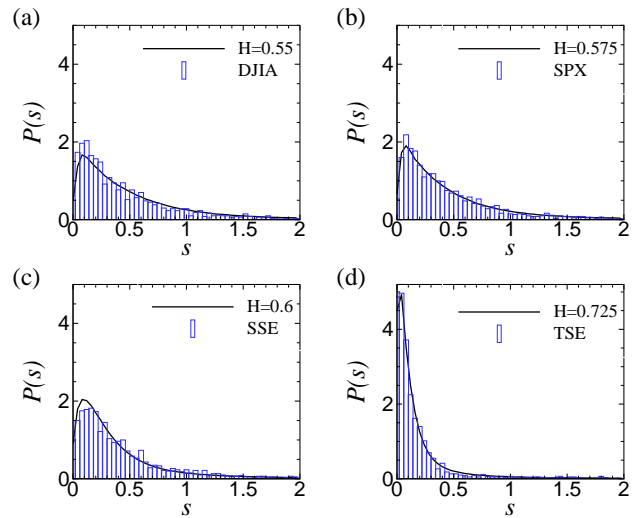}
\caption[NNES_finance]{(Color online) The NNES distribution of the normalized daily return of DJIA, SPX, SSE, and TSE shows a good agreement with that of the optimal Hurst exponent of the associated fGn family.}
\label{NNES_finance}
\end{figure}

\begin{figure*}[t]
\centering
\includegraphics[width=1\linewidth]{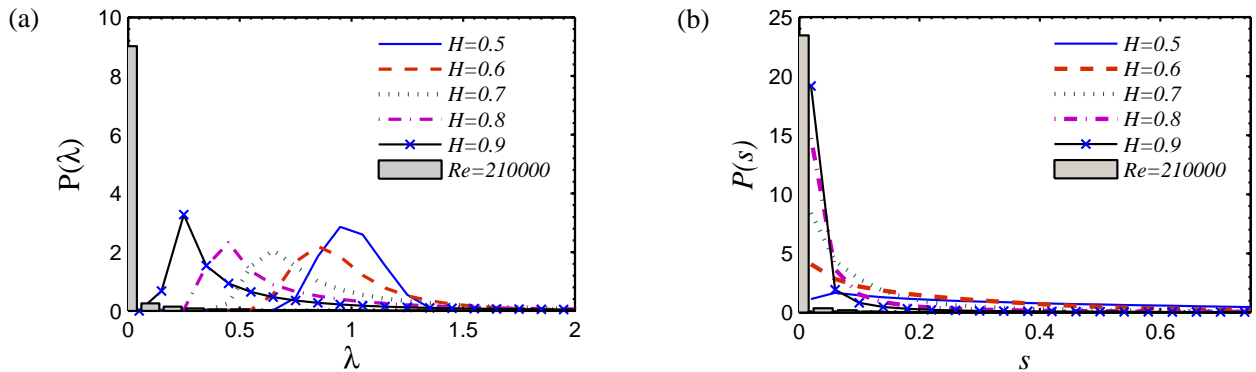}
\caption[ES_and_NNES_Velocity]{(Color online) (a) The ES and (b) the NNES distribution of the normalized turbulence velocity time series with $Re=210000$ compared to those of the fGns family with various values of the Hurst exponents. None of the used Hurst exponents provide a fair fit to the turbulence data implying the existence of non-fGn correlations in the turbulence phenomenon. One may note additionally that it is unnecessary to examine fGns with $H<0.5$ in the figure, since it is known that the velocity profile of turbulence exhibits correlations of positive nature comparable with the fGn family only with $H>0.5$.}
\label{ES_and_NNES_Velocity}
\end{figure*}

For financial time series we have used various actual databases covering securities from the Dow Jones Industrial Average (DJIA), the Standard \& Poor's 500 (SPX), the Shanghai Stock Exchange (SSE), and the Tehran Stock Exchange (TSE). We have aimed at analyzing the daily change of the DJIA and SPX from January 1, 1980 until January 1, 2011. The starting points for SSE and TSE are January 1, 2001 and January 1, 1996, respectively. In order to quantify the correlations, we first calculate the daily return time series of stock given by
\begin{eqnarray}
\label{eq:return}
R_t=\ln(S_{t+1})-\ln(S_t) \;,
\end{eqnarray}
where $S_t$ denotes the price at time $t$ of each of the four stocks considered above. Since each stock has a different standard deviation, we define a normalized return of the form
\begin{eqnarray}
\label{eq:normelized return}
r_t \equiv \frac{R_t-\langle R\rangle_{\mathrm{time}}}{\sigma} \; ,
\end{eqnarray}
where $\sigma =\sqrt{\langle R^2\rangle_{\mathrm{time}}-\langle R\rangle_{\mathrm{time}}^2}$ is the standard deviation of $R_t$. We construct then the autocorrelation matrices (associated with the four stock series) of a paradigmatic size $1500 \times 1500$ for the financial normalized return series so obtained. On the other hand, $M=500$ fGns associated with the Hurst exponents in some chosen positively correlated range $H=0.525,\ldots,0.8$ with the Hurst exponent increment $\delta H \equiv 0.025$ and the same length as that of the stock series are next generated. For each stock, the optimal Hurst exponent is found by searching the least relative error. Following the outlined recipe at the beginning of the current section, the errors of approximating $\mathbf{\Lambda}^0$ by $\overline{\mathbf{\Lambda}}$ for DJIA, SPX, SSE, and TSE read 0.040, 0.036, 11.25, and 0.094, respectively.

Figures~\ref{ES_finance} and~\ref{NNES_finance} illustrate the outlined operational prescription for finding the desired optimal Hurst exponents associated with each stock market. The results reveal a remarkable agreement with fGns in spite of the non-Gaussianity in returns of the stock market. On the other hand, since the distance from $H=0.5$ (associated with the while noise limit) can be regarded as a measure of the efficiency of a market, the method yields, at the same time, an operational recipe for telling apart an efficient market from an emerge one.

\subsection{Turbulence}
\label{sec:Turbulence}

\begin{figure*}[tb]
\centering
\includegraphics[width=1\linewidth]{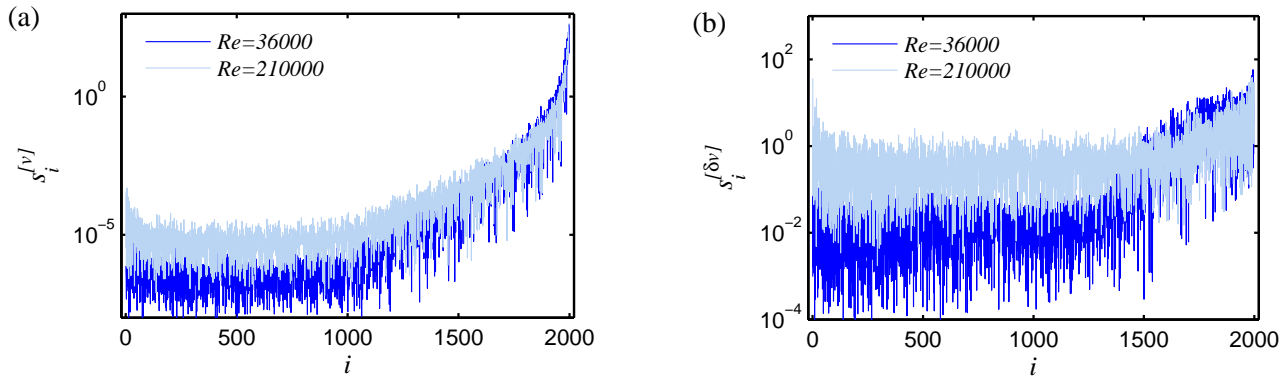}
\caption[NNES_of_turbulence]{(Color online) Adjacent eigenvalues spacing $s_i=\tilde{\lambda}_{i+1}-\tilde{\lambda_i}$, $i=1,2,\ldots,N$ for (a) velocity profile $v(t)$ and (b) velocity increment profile $\delta v(t)=v(t+1)-v(t)$ of low and high Reynold's numbers. The two Reynolds numbers display noticeably different trends in their adjacent eigenvalues spacing in contrast to their indistinguishable ES and NNES distribution.}
\label{NNES_of_turbulence}
\end{figure*}

Another important example belongs to the paradigm of the turbulence velocity profile $V(t)$ with two Reynold's numbers $Re = 36000$ and 210000. These time series indicate the local velocity measured at a fixed point in the turbulent region of a round free jet. In order to realize a fair comparison of such series with fGns and also with each other we invoke a normalization scheme as
\begin{eqnarray}
\label{eq:normelized return}
v(t)=\frac{V(t)-\langle V \rangle_{\mathrm{time}} }{\sigma}  \;,
\end{eqnarray}
where $\sigma=\sqrt{\langle V^2\rangle_{\mathrm{time}}-\langle V\rangle_{\mathrm{time}}^2}$ is the standard deviation of $V(t)$. We then proceed with obtaining the statistics of such a normalized $v(t)$ series similar to that of the financial time series. In this calculation, the size of the autocorrelation matrix is taken $N=2000$ and $M=500$ fGns series with the same length as the turbulence series are generated for each Hurst exponent in the range $H=0.5,\ldots,0.9$ with $\delta H = 0.1$.

Figure~\ref{ES_and_NNES_Velocity} contrasts the results of the calculation of the ES and NNES distribution associated with the high Reynold's number $Re = 210000$ to those of the fGns with various Hurst exponents. We point out the choice of the high Reynold's number for the purpose of this figure is fully arbitrary and one could equally consider the low number with almost the same ES and NNES distributions giving rise to an invisible distinction of the associated data (the bars in the plot) from those of the high Reynold's number. Quite strikingly, none of the Hurst exponents provide a good fit to the turbulence data. Unlike the financial correlations, this implies significant difference between the nature of turbulent correlations and those of fGn. The search for an alternative family of time series capable of capturing the turbulent correlations thus remains open.

Finally, as a by-product of the autocorrelation analysis of fGn, we have found out that in spite of the indistinguishability of the low and high Reynold's number in straight analysis of the \emph{distribution} of the eigenvalues as described above, a proper distinction between them becomes feasible upon analyzing instead their adjacent eigenvalues spacing $s_i=\tilde{\lambda}_{i+1}-\tilde{\lambda_i}$, $i=1,2,\ldots,N$ as illustrated on a semi-log scale in Fig.~\ref{NNES_of_turbulence}. The results for the adjacent eigenvalues spacing of (a) the velocity profile $s_i^{[v]}$ as well as (b) the velocity increment profile $s_i^{[\delta v]}$ with ${\delta v(t)} \equiv v(t+1)-v(t)$ reveal discernibly different trends for different Reynold's numbers.

\section{Concluding remarks}
\label{sec:conclusions}

In this work, we have analyzed the autocorrelation matrix of a time series using an RMT technique. For this purpose, it has been demonstrated that the fGns family provides an \emph{in situ} benchmark and figure of merit for accessing correlation information of the empirical systems in a way which is unattainable to a brute force RMT approach. In a nutshell, such information encompass the followings:

(i) The average of the eigenvalues of the fGn's autocorrelation matrix in the negatively correlated region is smaller than the one in the negatively correlated region.

(ii) The eigenvalues of the positively correlated series (associated with the higher values of the Hurst exponents) tend to attract each other whereas the negatively correlated ones (associated with the lower values of the Hurst exponents) rather show a tendency to repel each other.

(iii) The number of significant participants in the eigenvector associated with the largest eigenvalue proves larger in the positively correlated series compared to the one in the negatively correlated region.

(iv) In the first context of the financial time series, it appeared that although the return PDF of the stock market is known to be non-Gaussian, but its correlation content exhibits a good agreement with an fGn. This, in turn, promises to provide a powerful tool for distinguishing an efficient market from the emerge one.

(v) In the context of the turbulence, our results suggest a significant discrepancy from fGns in spite of a Gaussian velocity profile assumed in the description of the phenomenon. Nonetheless, our approach provides a systematic recipe for distinguishing various Reynold's numbers.

\bibliography{RMT_fgn}

\end{document}